\documentclass[a4paper,11pt]{article}
\usepackage{jheppub}
\usepackage[T1]{fontenc}
\usepackage{amsmath,amssymb,amsfonts}
\usepackage{bm}
\usepackage{graphicx}
\usepackage{hyperref}
\usepackage{booktabs}
\usepackage{orcidlink}
\usepackage{enumitem}
\usepackage{physics}
\usepackage{caption}
\usepackage{tikz}
\usetikzlibrary{decorations.pathmorphing,decorations.markings,arrows.meta}
\usepackage{subcaption}
\usepackage{float}
\usepackage[colorinlistoftodos]{todonotes}
\newcommand{\ii}{\mathrm i}
\newcommand{\Lie}{\mathcal L}
\newcommand{\barg}{\bar g}
\newcommand{\calR}{\mathcal R}

\newcommand{\cD}{\mathcal D}

\title{\boldmath Gravitational Memory from AdS States in the Flat-Space Limit}

\author[a]{David A. Lowe \orcidlink{0000-0002-3168-3612}}
\author[a]{Juanyi Yang \orcidlink{0009-0007-8425-4612}}
\affiliation[a]{Department of Physics, Brown University, Providence, RI, 02912, USA}
\emailAdd{lowe@brown.edu}
\emailAdd{juanyi\_yang@brown.edu}

\abstract{
We study linearized gravitational reconstruction in global AdS$_4$ with dynamical scalar matter, building on vacuum Einstein perturbations with fixed boundary metric and nontrivial normalizable data. External scalar sources prepare an excited state and modify the holographic Ward identities, while subsequent evolution can proceed without boundary driving. Separating the bulk matter-sourced response from vacuum perturbations, we derive an electric quadrupole master equation with the matter-dependent corrections required by the Einstein constraints. We also verify that a cylindrical boundary integral reproduces the Regge--Wheeler master field mode by mode, without identifying it with the full global Fefferman--Graham reconstruction. For weakly interacting scalars that become massless in the flat-space limit, we extract the Bondi shear and news from retarded gravitational evolution. The resulting supertranslation memory agrees numerically with the outgoing energy redistribution and the quadrupole moment of the boundary work. In the ordered flat-space limit, later boundary controls leave the first event's waveform unchanged, despite cancelling residual gravitational modes at finite AdS radius.
}
\begin{document}
\maketitle

\flushbottom
\section{Introduction}
\label{sec:intro}

A basic test of the AdS/CFT correspondence \cite{Maldacena,GKP,Witten} is the explicit recovery of bulk fields from boundary observables. In the gravitational setting, the boundary stress tensor specifies a normalizable linearized metric response only after the boundary metric representative, bulk regularity conditions, and gauge have been fixed. While explicit smearing representations were first developed for matter fields in \cite{BDHM,HKLL,HKLLcomplex,KLL}, extending bulk reconstruction to gravity requires careful treatment of both gauge invariance and gravitational dressing \cite{DonnellyGiddings}.

The reconstruction of linearized gravitational perturbations in the absence of bulk matter sources is provided by the global framework of \cite{LWYGlobalReconstruction}. These perturbations solve the vacuum Einstein equations but may carry nonzero normalizable boundary stress-tensor data, even though the boundary metric is held fixed. The global framework constructs regular scalar- and vector-type gravitational normal modes on AdS$_4$, fixes their holographic stress-tensor normalization, and establishes their boundary-anchored Fefferman--Graham (FG) gauge improvements. Importantly, it also distinguishes local light-cone-cut formulas from the reconstruction of the full global FG metric, showing that radial gauge-restoring data need not be supported on the nearest cuts even for smooth normal-mode configurations. We adopt such global spectral reconstruction as our homogeneous input, where "homogeneous" refers to the absence of a bulk matter source in the linearized Einstein equations. It does not imply vanishing gravitational perturbations or boundary stress-tensor data, and with the Dirichlet boundary metric fixed, the normalizable gravitational response can still be nonzero and is reconstructed by the vacuum spectral kernel. As a supplementary mode-level check, we also verify directly in the quadrupole sector that the master field extracted from the cylindrical boundary integral of \cite{LWY} matches after transforming to Regge--Wheeler (RW) gauge, though this local check does not substitute for the full global FG improvement.

The central question of the present paper is how this picture changes when the bulk metric is sourced by a dynamical scattering process. The full holographic stress tensor includes the matter response and generally satisfies inhomogeneous Ward identities when scalar boundary sources are active. It cannot therefore be identified with the homogeneous vacuum reconstruction datum. To overcome this, we separate a particular solution of the sourced Einstein equations from the homogeneous gravitational response and identify the exact subtracted boundary tensor on which the vacuum reconstruction acts. We then reduce the sourced electric quadrupole equations to a single radial master equation while retaining all source-dependent terms in the metric reconstruction map and the zero-frequency constraints. Our master-variable reduction builds on the classic formalism of Regge, Wheeler, Zerilli, and Moncrief \cite{ReggeWheeler,Zerilli,Moncrief}, its higher-dimensional extension by Kodama and Ishibashi \cite{KI}, and analyses of AdS gravitational boundary conditions \cite{IW,DiasSantos}.

We apply this sourced framework to the gravitational memory effect, in which an outgoing radiation pulse leaves a permanent shift in the asymptotic shear governed in Bondi gauge by a supertranslation \cite{ZeldovichPolnarev,Christodoulou,Favata,Bondi,Sachs,Weinberg,Strominger,HeLysovMitraStrominger,SZ}. As a concrete realization, we consider two weakly interacting scalar fields that undergo a smooth scattering process with a nonzero quadrupolar redistribution of energy. We solve for the resulting sourced gravitational waveform, transform it to Bondi gauge in a controlled flat-space limit, and compare its asymptotic memory offset against both the outgoing scalar flux and an angular moment of the work performed at the AdS boundary.

The scalar boundary sources prepare the incoming state, but they are not needed to sustain the subsequent matter interactions. When the sources are switched off, the excited state evolves under the undeformed CFT Hamiltonian while its bulk matter stress continues to drive gravitational perturbations. In the ordered flat-space limit, the later boundary controls and delayed gravitational completion lie outside the causal observation region of the first scattering event. The Bondi shear and news extracted from this event therefore agree with those obtained without the later sources.

Related aspects of asymptotic symmetries in AdS have been studied using impulsive superrotation geometries \cite{LoweRamirez} and the $\Lambda$-BMS dictionary relating the Bondi and FG descriptions \cite{CompereFiorucciRuzziconi}. Whereas the latter approach allows the boundary metric to fluctuate, we hold the cylindrical boundary representative fixed throughout. In doing so, we carefully distinguish the finite-$L$ gravitational completion and its late-time behavior on the cylinder from the flat-space limit of the initial scattering event.

Regarding the scope of our methods, the homogeneous mode normalizations and gauge conventions imported from \cite{LWYGlobalReconstruction} are purely analytic, as are our new sourced master reduction and asymptotic gauge transformations. On the other hand, the dynamical waveform for our scalar scattering model and its quantitative comparison with the matter flux rely on numerical time evolution, and we do not claim a uniform flat-space limit across arbitrary source histories.

The remainder of the paper is organized as follows. Section \ref{sec:setup} formulates the source subtraction and identifies the conserved boundary datum on which the vacuum reconstruction acts. Section \ref{sec:kernel} summarizes the homogeneous global input alongside our quadrupole consistency check. Section \ref{sec:source} develops the sourced gravitational inverse, and section \ref{sec:memory} analyzes the scalar scattering process and its relative supertranslation memory. Technical derivations covering the vacuum consistency check, the sourced reduction, the scalar evolution, and the waveform extraction are collected in the appendices.

\section{Boundary data and source subtraction}
\label{sec:setup}

We work in dimensionless global coordinates on AdS$_4$ with background metric
\begin{equation}
 L^{-2}d\bar s^2=-Fd\tau^2+F^{-1}d r^2+r^2q_{AB}d x^Ad x^B\,,
 \qquad F=1+r^2\,,\qquad r=\tan\rho\,,
 \label{eq:background}
\end{equation}
where $0\leq\rho<\pi/2$ and $q_{AB}$ is the metric on the unit round sphere. The physical time and radial coordinates are given by $t=L\tau$ and $R=Lr$, and we choose the cylindrical boundary representative $\gamma_{ij}d x^id x^j=-d\tau^2+q_{AB}d x^Ad x^B$. Throughout the paper, Greek indices label bulk directions, $i,j$ label boundary directions, and $A,B$ label angles on the sphere. In the radial gravitational formulas below we set $L=1$ except where $L$ is displayed explicitly.

In the presence of a conserved matter stress tensor $S_{\mu\nu}$, the linearized Einstein equations read
\begin{equation}
 \mathcal E[h]_{\mu\nu}\equiv\delta R_{\mu\nu}[h]+\frac3{L^2}h_{\mu\nu}
 =\calR_{\mu\nu}\,,\qquad
 \calR_{\mu\nu}=8\pi G_4\left(S_{\mu\nu}-\frac12\barg_{\mu\nu}S\right)\,.
 \label{eq:einstein}
\end{equation}
Given a particular solution $h_P$ with prescribed initial and boundary data, the difference $h-h_P$ satisfies the homogeneous linearized equations. Although this subtraction is algebraically elementary, it is physically meaningful only when the two boundary responses are compared in the same boundary frame and for the exact same matter evolution.

In FG coordinates the bulk metric is $L^2z^{-2}(d z^2+g_{ij}d x^id x^j)$, and its near-boundary expansion is $g_{ij}=\gamma_{ij}+z^2g_{(2)ij}+z^3g_{(3)ij}+\cdots$.
Varying the bulk action supplemented by standard boundary counterterms yields the renormalized holographic stress tensor \cite{BalasubramanianKraus,deHaro}. For the scalar fields considered below, we adopt standard quantization with near-boundary falloff $\phi_I=zJ_I+z^2V_I+\cdots$, and we include the canonical scalar counterterm $-(2L)^{-1}\int\sqrt{-\gamma_{\rm cut}}\sum_I\phi_I^2$ without additional finite counterterms. In this renormalization scheme, the boundary stress tensor and scalar expectation values take the form
\begin{equation}
 T_{ij}=\frac{3L^2}{16\pi G_4}g_{(3)ij}
        +L^2\sum_IJ_IV_I\gamma_{ij}\,,\qquad
 \langle O_I\rangle=L^2V_I\,.
 \label{eq:stress}
\end{equation}
Here the contribution proportional to $J_IV_I$ is a scalar contact term. The full stress tensor obeys the diffeomorphism and trace Ward identities $\nabla^iT_{ij}=\sum_I\langle O_I\rangle\partial_jJ_I$ and $T^i{}_i=\sum_IJ_I\langle O_I\rangle$ \cite{BianchiFreedmanSkenderis,Skenderis}. These identities are generally inhomogeneous while the external scalar sources are active. On any interval where $J_I=0$, they reduce to
\begin{equation}
 \nabla^iT_{ij}=0\,,\qquad T^i{}_i=0\,.
 \label{eq:sourcefreeward}
\end{equation}
The homogeneous Ward identities hold even when the scalar expectation values remain nonzero. They do not imply that the bulk matter stress $S_{\mu\nu}$ vanishes or that the gravitational perturbation obeys vacuum Einstein equations.

Crucially, these contact and source terms cancel in the difference $T_{ij}-T_{P,ij}$ provided $h$ and $h_P$ share the complete bulk matter solution rather than merely the boundary sources $J_I$. Indeed, two scalar configurations with identical sources $J_I$ can still differ in their normalizable responses $V_I$, which would alter both the bulk source $\calR_{\mu\nu}$ and the boundary contact term. Matching the full matter state isolates the conserved and traceless response difference
\begin{equation}
 \widehat t_{ij}=g_{(3)ij}-g_{(3)P,ij}
 =\frac{16\pi G_4}{3L^2}(T_{ij}-T_{P,ij})\,,\qquad
 \nabla^i\widehat t_{ij}=0\,,\qquad \gamma^{ij}\widehat t_{ij}=0\,.
 \label{eq:subtraction}
\end{equation}

With the shared matter history and fixed boundary representative in place, the full metric reconstruction takes the form
\begin{equation}
 h_{\rm rec}=h_P+L^2\mathcal K^{\rm FG}_{\rm glob}[\widehat t]
       +\Lie_\zeta\barg\,.
 \label{eq:reconstruction}
\end{equation}
Here $\mathcal K^{\rm FG}_{\rm glob}$ denotes the global spectral vacuum reconstruction operator of \cite{LWYGlobalReconstruction} restricted to the electric quadrupole sector, rather than a local distribution supported solely on global light-cone cuts, and it acts exclusively on the subtracted response $\widehat t_{ij}$ rather than on the full sourced stress tensor $T_{ij}$. Finally, the Lie derivative $\Lie_\zeta\barg$ serves purely to align the bulk gauges in which the particular and homogeneous solutions are presented, vanishing identically whenever both contributions are already expressed in the same boundary-anchored FG gauge.

\section{Homogeneous input from global AdS reconstruction}
\label{sec:kernel}

\subsection{Boundary harmonics and the physical radial modes}

We summarize the vacuum data required below, using the global normal-mode construction and FG normalization of \cite{LWYGlobalReconstruction}. The complete local conformal-frame cut construction and the boundary-anchored global gauge improvement are given there and will not be repeated here. For the electric quadrupole, let $Y=Y_{2m}$ and $Y_{AB}=D_AD_BY+3q_{AB}Y$. Conservation and tracelessness on the cylinder fix all components of the boundary tensor in terms of a single amplitude:
\begin{equation}
 \left(t_{\tau\tau},t_{\tau A},t_{AB}\right)
 =a\left(4Y,\frac{2\ii\omega}{3}D_AY,
  2q_{AB}Y+\frac{3-\omega^2}{3}Y_{AB}\right)e^{-\ii\omega\tau}\,.
 \label{eq:ttmode}
\end{equation}
This is the $\ell=2$ specialization of the scalar-type Ward solution in \cite{LWYGlobalReconstruction}. The coefficient $a$ is one quarter of the energy-density harmonic coefficient, rather than an independent tensor-channel normalization.

For linearized vacuum Einstein perturbations about global AdS with a fixed boundary metric, the regular master function satisfies
\begin{equation}
 \left(-\partial_\rho^2+\frac{6}{\sin^2\rho}\right)p=\omega^2p,
 \qquad p'(\pi/2)=0\,,
 \label{eq:mastereqn}
\end{equation}
where regularity at $\rho=0$ selects the solution behaving as $\rho^3$. The resulting frequencies are $\omega_n=3+2n$ and a convenient boundary-normalized basis is
\begin{equation}
 e_n^{(b)}(\rho)=
 \frac{\sin^3\rho\,C_{2n}^{(3)}(\cos\rho)}{C_{2n}^{(3)}(0)},
 \qquad C_{2n}^{(3)}(0)=(-1)^n\frac{(n+1)(n+2)}2,
 \qquad e_n^{(b)}(\pi/2)=1\,.
 \label{eq:gravmodes}
\end{equation}
These are the $\ell=2$ form of the hypergeometric scalar master modes of \cite{LWYGlobalReconstruction}. In particular, the unit-density mode $t_{\tau\tau}=Ye^{-\ii\omega_n\tau}$ corresponds to $p=e_n^{(b)}/4$. Thus the boundary-stress normalization and the radial normalization in our subsequent sourced equations coincide without an additional rescaling.

For later use we retain the RW differential map from $p$ to the metric:
\begin{align}
 L^{-2}h^{\rm RW}_{\mu\nu}d x^\mu d x^\nu
 &=e^{-\ii\omega\tau}\left[AYd\tau^2+2BYd\tau d r
   +\frac{A}{F^2}Yd r^2+r^2K Yd\Omega_2^2\right],
       \label{eq:rwmetric}\\
 K&=Fp_{,r}+\frac{3p}{r},\qquad
 B=-\ii\omega\left(rp_{,r}+\frac pF\right),\nonumber\\
 A&=Fp_{,r}+\left(\frac{3F}{r}-r\omega^2\right)p\,.
 \label{eq:Dmap}
\end{align}
In eqs. \eqref{eq:rwmetric}--\eqref{eq:Dmap}, $p$ denotes the radial amplitude with the common $e^{-\ii\omega\tau}Y$ suppressed. We call this differential map $\cD$. The RW representative is convenient for the sourced Einstein equations. Its relation to the boundary-anchored FG representative is fixed by the vacuum reconstruction of \cite{LWYGlobalReconstruction}.

\subsection{Spectral reconstruction and its domain}
For an electric quadrupole response supported on the allowed normal-mode frequencies, write
\begin{equation}
 t_{\tau\tau}
 =4\sum_{n,m,\pm}a_{nm}^{\pm}Y_{2m}e^{\mp\ii\omega_n\tau},
 \qquad
 \mathcal B_2[t]
 =\sum_{n,m,\pm}a_{nm}^{\pm}e_n^{(b)}Y_{2m}
           e^{\mp\ii\omega_n\tau}\,.
 \label{eq:modepairing}
\end{equation}
The other components of $t$ are those fixed by eq. \eqref{eq:ttmode}. For finite sums, or for smooth mode sums obeying
\begin{equation}
 \sum_{n,m,\pm}(1+\omega_n)^N|a_{nm}^{\pm}|<\infty
 \quad\text{for every }N\,,
 \label{eq:rapid}
\end{equation}
the global spectral mode construction gives, in the quadrupole sector,
\begin{equation}
 L^2\mathcal K^{\rm FG}_{\rm glob}[t]
 =\cD\mathcal B_2[t]+\Lie_{\xi_{\rm FG}[t]}\barg\,,
 \label{eq:theorem}
\end{equation}
with the boundary-anchored gauge transformation specified in
\cite{LWYGlobalReconstruction}. Equation \eqref{eq:theorem} is used here as a
known vacuum result. In particular, the global FG improvement is not in general
represented by an ordinary distribution on the nearest light-cone cuts.

A separate consistency check concerns the cylindrical boundary integral $\mathcal K$
associated with \cite{LWY}, defined in appendix \ref{app:intcheck}. For a single quadrupole normal mode with
$t_{\tau\tau}=Y_{2m}e^{-\ii\omega_n\tau}$, direct integration followed by a
transformation to RW gauge gives
\begin{equation}
 p_{\rm int}(\rho)
 =\frac12\left(rK_{\rm RW}-\frac{\ii F}{\omega_n}B_{\rm RW}\right)
 =\frac14e_n^{(b)}(\rho),\qquad \omega_n=3+2n\,.
 \label{eq:intmastercheck}
\end{equation}
The coefficients $B_{\rm RW}$ and $K_{\rm RW}$ are read from the integral after
transforming to RW gauge, rather than from its original gauge. The remaining RW
coefficients satisfy eq. \eqref{eq:Dmap} with the same $p_{\rm int}$.
Appendix \ref{app:intcheck} records the integral prescription and the reduced
calculation. This tests the vacuum metric up to a diffeomorphism for each
normal mode. It does not identify the integral with the complete global FG
kernel, which requires the boundary-anchored improvement of
\cite{LWYGlobalReconstruction}.

The mode sum in eq. \eqref{eq:modepairing} is well defined on compact regular
bulk regions under the decay assumption \eqref{eq:rapid}. We do not infer a
uniform flat-space limit or a reconstruction of arbitrary distributional data.
For the retarded matter process considered below, the physical gravitational
response is contained in $h_P$ unless independent homogeneous data are added.

\section{The gravitational inverse with a matter source}
\label{sec:source}

Once a particular solution $h_P$ is known, the spectral reconstruction of the preceding section supplies any freely specified homogeneous contribution without applying a vacuum boundary kernel to the matter-dependent stress tensor. This subtraction remains necessary during source-free intervals, since a conserved and traceless total boundary stress tensor can still include the response to nonvanishing bulk matter. Here we construct $h_P$ directly from the sourced Einstein equations for the source class required in section \ref{sec:memory}, namely an electric quadrupole generated by a spherically symmetric scalar background together with a linear quadrupole perturbation. At this perturbative order, the traceless angular component of $\calR_{AB}$ vanishes identically. Throughout this section we set $L=1$ and take all fields to have harmonic time dependence $e^{-\ii\omega\tau}$.

Decomposing the source components as $\calR_{\tau r}=R_sY$, $\calR_{\tau A}=U_sD_AY$, $\calR_{rA}=V_sD_AY$, and $\calR_{AB}=Q_sq_{AB}Y$, the matter Bianchi identity reads $\bar\nabla^\mu\calR_{\mu\nu}-\tfrac12\partial_\nu\calR^\mu{}_{\mu}=0$. Eliminating the constraints reduces the six independent Einstein equations to the single sourced master equation
\begin{equation}
 \left(\partial_\rho^2+\omega^2-\frac{6}{\sin^2\rho}\right)p_P=\mathcal S_{\rm eff}\,.
 \label{eq:sourcedmaster}
\end{equation}
The bulk metric is then recovered from eq. \eqref{eq:Dmap} supplemented by the explicit source shifts
\begin{equation}
 K\mapsto K+\kappa\,,\qquad
 B\mapsto B-\frac{\ii\omega r\kappa}{F}\,,\qquad A\mapsto A+\alpha_s\,,
 \label{eq:sourcecorrections}
\end{equation}
where the expressions for $\kappa$, $\alpha_s$, and $\mathcal S_{\rm eff}$ are recorded in appendix \ref{app:source}. Retaining these source corrections is essential, since inserting a solution of the sourced master equation directly into the unmodified vacuum metric map would violate the Einstein constraints.

The scattering process analyzed below obeys the homogeneous Neumann boundary condition $\partial_\rho p_P(b)=0$ at $b=\pi/2$. Introducing the center-regular homogeneous solution $u_\omega\sim\rho^3$ and the boundary-normalized solution $v_\omega$ satisfying $v_\omega(b)=1$ and $v'_\omega(b)=0$, the Wronskian and radial Green's function take the form
\begin{equation}
 W_\omega=\frac{15\cos(\pi\omega/2)}{\omega^2-1}\,,\qquad
 G_\omega(\rho,\rho')=
 \frac{u_\omega(\rho_<)v_\omega(\rho_>)}{W_\omega}\,,\qquad
 p_P(\rho)=\int_0^bG_\omega(\rho,\rho')\mathcal S_{\rm eff}(\rho')\,d\rho'\,.
 \label{eq:green}
\end{equation}
We select the retarded response via the contour prescription $\omega\to\omega+\ii0$, which simultaneously regulates the inverse powers of $\omega$ generated during the constraint elimination. While eq. \eqref{eq:green} relies on the homogeneous Neumann condition $\partial_\rho p_P(b)=0$, more general matter boundary conditions can be accommodated by including an inhomogeneous radial boundary lift.

At finite $L$, the residual normal-mode ringing excited by the scattering event can be cancelled cleanly by repeating the identical stress-energy history with an independent copy of the matter fields after a dimensionless time delay $\Delta\tau=\pi$. Because the two sources add linearly in the gravitational sector, the combined prefactor
\begin{equation}
 \frac{1+e^{\ii\pi\omega}}{W_\omega}
 =\frac{2e^{\ii\pi\omega/2}(\omega^2-1)}{15}\,.
 \label{eq:echo}
\end{equation}
is entire, removing all of the quadrupole normal-mode poles. In the time domain, each oscillatory sine mode in the retarded spectral Green's function is cancelled by its delayed counterpart once the second event concludes. Potential zero-frequency poles arising from the constraint elimination require separate inspection, and we show in appendix \ref{app:source} that stress-energy conservation cancels their apparent residues in the metric. Because the spatial operator $-\partial_\rho^2+6/\sin^2\rho$ admits no regular static Neumann mode, the metric perturbation remaining after both compactly supported matter processes have terminated is pure gauge within the fixed boundary frame. This cancellation is a linearized statement within the electric quadrupole sector and does not imply an exact nonlinear return to pure AdS spacetime.

We verify this sourced reconstruction both algebraically and numerically. As shown in appendix \ref{app:source}, after imposing the master equation \eqref{eq:sourcedmaster}, substituting the reconstructed metric together with the source corrections \eqref{eq:sourcecorrections} into the covariant linearized Einstein equations reduces all components identically to the matter conservation laws. In addition, an independent numerical comparison between the radial Green's function integral and a direct boundary-value solver provides a quantitative check of the radial inverse.

\section{A scalar process with relative supertranslation memory}
\label{sec:memory}
\subsection{Matter pulse and boundary control}
Consider two minimally coupled scalars with action
\begin{equation}
 I=\frac1{16\pi G_4}\int\sqrt{-g}\,(R+6/L^2)
 -\int\sqrt{-g}\left[
 \frac12\sum_{I=1}^2\bigl((\partial\phi_I)^2-2\phi_I^2/L^2\bigr)
 +\frac g4\phi_1^2\phi_2^2\right]\,.
 \label{eq:action}
\end{equation}
The scalar fields have $m^2L^2=-2$ and are treated in standard quantization, with dual operator dimension $\Delta=2$. Since the physical mass satisfies $m^2=-2/L^2$, keeping $\Delta$ fixed as $L\to\infty$ gives massless scalar fields in flat spacetime. A finite nonzero flat-space mass would instead require a simultaneous large-$L$, large-$\Delta$ limit with $\Delta/L$ held fixed. Such a limit involves a different matter spectrum and scattering problem, which we do not consider here. 

We retain leading gravitational backreaction, first order in $g$, and first order in an anisotropy $\varepsilon$. With $Y=P_2(\cos\theta)$ and $\psi_I=R\phi_I$, choose equal species,
\begin{equation}
 \psi_I=\psi_0^{(0)}+\varepsilon\psi_2^{(0)}Y
     +g\bigl(\psi_0^{(1)}+\varepsilon\psi_2^{(1)}Y\bigr)+\cdots\,.
 \label{eq:matterexp}
\end{equation}
Two incoming generators $A_0$ and $A_2$ determine the regular free waves. The first Born corrections satisfy
\begin{equation}
 (\partial_\tau^2+H_0)\psi_0^{(1)}=-\frac{(\psi_0^{(0)})^3}{2\sin^2\rho},
 \qquad
 (\partial_\tau^2+H_2)\psi_2^{(1)}=-\frac{3(\psi_0^{(0)})^2\psi_2^{(0)}}{2\sin^2\rho}\,,
 \label{eq:bornads}
\end{equation}
where $H_\ell=-\partial_\rho^2+\ell(\ell+1)/\sin^2\rho$. We first solve these equations with zero incoming correction and homogeneous scalar Dirichlet data. The resulting Born field therefore has reflected residual modes, even when the free pulse has gone. However, those modes can be removed by an explicit boundary source. Their frequencies are the scalar Dirichlet frequencies $\omega_{\ell n}=\ell+2+2n$, not the gravitational frequencies of section \ref{sec:kernel}. After the free pulse has vanished, let $(q_n,p_n)$ be the residual modal position and velocity, and let $d_n=e'_{\ell n}(b)$ for an orthonormal scalar mode. A smooth boundary control is
\begin{equation}
 j_\ell(s)=\frac2\pi w(s)\sum_n
 \frac{p_n\cos\omega_{\ell n}s-\omega_{\ell n}q_n\sin\omega_{\ell n}s}{d_n},
 \qquad \sum_{k\in\mathbb Z}w(s+k\pi)=1\,,
 \label{eq:control}
\end{equation}
Here $w$ has support in $[0,2\pi]$, and its Fourier moments cancel the residual scalar modes. The initial boundary sources prepare incoming scalar waves, which then propagate and interact in the bulk without continued boundary driving. For appropriately localized compactly supported generators, the sources vanish during the first scattering event. The control is applied only after this event and after the free pulse has vanished. Its contribution to the stress tensor begins at order $g^2$ and therefore does not change the $\mathcal O(g\varepsilon)$ energy redistribution considered here. This construction is exact for smooth compactly supported generators with compatible data. For the Gaussian profiles used below, the separation holds to the accuracy at which their small tails are neglected. The free fields and the moment calculation are given in appendix \ref{app:matter}.

\subsection{Boundary work and the memory observable}

The natural boundary observable governing the energy transfer during this scattering process is the full density of injected power,
\begin{equation}
 \mathcal P\equiv\nabla_iT^{i\tau}
 =-\sum_I\langle O_I\rangle\partial_\tau J_I
 =\sum_I\psi_{I,\rho}(\tau,b)\psi_{I,\tau}(\tau,b)\,,
 \label{eq:power}
\end{equation}
where the second equality uses $J_I=\psi_I(b)/L$ and $V_I=-\psi_{I,\rho}(b)/L$. Note that eq. \eqref{eq:power} requires the complete sourced stress tensor \eqref{eq:stress}, whereas the transverse-traceless response difference \eqref{eq:subtraction} isolates the homogeneous gravitational degrees of freedom and cannot substitute for $T_{ij}$ in the work integral.

In the flat-space limit, we describe future null infinity using the Bondi expansion \cite{Bondi,Sachs}, with $g_{AB}=R^2q_{AB}+RC_{AB}+\cdots$ and $g_{uu}=-1+2m_B/R+\cdots$. Electric-parity memory is generated by a relative supertranslation $f$ according to \cite{SZ}
\begin{align}
 \Delta C_{AB}&=-2\left(D_AD_B-\tfrac12q_{AB}D^2\right)f\,,\label{eq:memorydef}\\
 D^2(D^2+2)f&=-4\left[\Delta m_B+4\pi G_4\mathcal E_+\right]_{\ell\geq2}\,,
 \label{eq:bondimemory}
\end{align}
where $\mathcal E_+=\int d u\,[T_{uu}^{(0)}+N_{AB}N^{AB}/(32\pi G_4)]$ and $N_{AB}=\partial_uC_{AB}$. Equation \eqref{eq:bondimemory} follows from the integrated Bondi mass-loss constraint in the conventions of appendix \ref{app:matter}, with the $\ell=0,1$ harmonics of $f$ fixed to zero. This relative BMS transformation between early and late retarded times is distinct from the bulk gauge improvement relating the vacuum Regge--Wheeler modes to the boundary-anchored Fefferman--Graham description of \cite{LWYGlobalReconstruction}.

Because our massless scattering process leaves behind no massive remnant, we can compare the outgoing radiation directly against the antipodally matched incoming hard distribution in a specified incoming soft frame \cite{Strominger,HeLysovMitraStrominger}. Under this incoming matching and endpoint condition, rather than as a general rule for omitting $\Delta m_B$ from eq. \eqref{eq:bondimemory}, the leading-order relation in $G_4$ simplifies to
\begin{equation}
 D^2(D^2+2)f=-16\pi G_4(\mathcal E_{\rm out}-\mathcal E_{\rm in}^{\rm matched})_{\ell\geq2}\,.
 \label{eq:relative}
\end{equation}
Denoting the quadrupole coefficient of this energy redistribution by $\Delta\mathcal E_2$, the supertranslation profile and memory amplitude take the form
\begin{equation}
 f=a_{\rm mem}P_2(\cos\theta)\,,\qquad
 a_{\rm mem}=-\frac{2\pi G_4}{3}\Delta\mathcal E_2\,.
 \label{eq:amemory}
\end{equation}
Using the normalization $\int P_2^2\,d\Omega=4\pi/5$, we can express the same memory amplitude in terms of the integrated boundary work as
\begin{equation}
 a_{\rm mem}=\lim_{L\to\infty}\frac{5G_4}{6L}\mathcal Q_2(L)\,,
 \qquad
 \mathcal Q_2(L)=\int_{\text{first process}}d\tau  d\Omega\,P_2\mathcal P
 \quad\text{at} \ \mathcal{O}(g\varepsilon)\,.
 \label{eq:workmemory}
\end{equation}

The boundary work records the energy exchanged with the external sources during the first preparation and scattering protocol. It need not represent continuing boundary driving while the scalar waves interact in the bulk. The time-dependent gravitational waveform is instead determined by the retarded Einstein equations with the bulk scalar stress tensor as their source. Equation \eqref{eq:workmemory} relates the energy transferred through the boundary to the resulting memory offset.

The integral in eq. \eqref{eq:workmemory} is restricted to the first process. At finite AdS radius, the later scalar controls and the gravitational completion contribute to the full global evolution. The latter repeats the matter history after a dimensionless delay $\pi$, corresponding to a physical delay $\pi L$. The evolution without the later boundary sources starts after preparation from the same complete scalar and gravitational Cauchy data, including the Born corrections and gravitational constraint data. In the ordered flat-space limit, we first take $L\to\infty$ with the initial scattering event and a fixed retarded-time observation region held fixed. The subsequent operations then lie outside the causal domain relevant to the retarded response of the first event. Thus the Bondi shear $C_{AB}$ and news $N_{AB}$ in this limit agree with those computed without the later boundary sources, in the same incoming Bondi frame. The statement is exact for the idealized compactly supported protocol and applies to the Gaussian example up to its neglected tails. Taking the late-time limit at fixed $L$ instead includes the delayed completion, so the two limits do not commute. In particular, a final quadrupole that is pure gauge at finite $L$ need not erase the memory extracted from the first event.

\subsection{A quantitative example}

To illustrate this relation quantitatively, we choose the incoming pulse profiles
\begin{equation}
 A_0(v)=\alpha e^{-(v/\sigma)^2}\,,\qquad
 A_2(v)=\alpha\sigma^2e^{-[(v-0.7\sigma)/\sigma]^2}\,,
 \label{eq:generators}
\end{equation}
working in the perturbative regime $g\alpha^2\ll1$, $G_4\alpha^2/\sigma^2\ll1$, and $|\varepsilon|\ll1$. Evaluating the retarded Born integral directly in flat spacetime yields
\begin{equation}
 \frac{\sigma\,\Delta\mathcal E_2}{g\varepsilon\alpha^4}\simeq4.3250\,,
 \qquad
 a_{\rm mem}\simeq-9.0583\,\frac{G_4g\varepsilon\alpha^4}{\sigma}\,.
 \label{eq:number}
\end{equation}
As shown in table \ref{tab:energycheck}, extrapolating the finite-$L$ boundary work reproduces this quadrupole energy redistribution. Specifically, fitting the finite-$L$ data to $c_0+c_2/L^2+c_4/L^4$ gives $c_0\simeq4.3250$ in the displayed units, in close agreement with the independent flat-space integral. Furthermore, setting the relative time shift between $A_0$ and $A_2$ to zero causes the quadrupole redistribution to vanish within numerical precision, while reversing the sign of the shift reverses the sign of $\Delta\mathcal E_2$ and leaves the monopole energy change consistent with zero. Together, these checks confirm that the memory offset arises from a genuine angular redistribution of energy rather than a net change in total energy, with the quoted values representing numerical estimates for the specific profiles in eq. \eqref{eq:generators} rather than universal constants or rigorous error bounds.

\begin{table}[tb]
\centering
\begin{tabular}{llr}
\toprule
Calculation & Cutoff or AdS radius & Normalized redistribution \\
\midrule
Flat retarded integral & $R_{\max}/\sigma=192$ & $4.325015$ \\
Boundary work & $L/\sigma=8$ & $4.380354$ \\
Boundary work & $L/\sigma=16$ & $4.339021$ \\
Boundary work & $L/\sigma=32$ & $4.328516$ \\
\bottomrule
\end{tabular}
\caption{Quadrupole energy redistribution for the profiles in eq. \eqref{eq:generators} with $\alpha=\sigma=1$. The flat-space entry reports $\Delta\mathcal E_2/(g\varepsilon)$, while the boundary-work entries report $-5\mathcal Q_2/(4\pi Lg\varepsilon)$. The three finite-$L$ values are used for the polynomial extrapolation in $L^{-2}$ described in the main text, and all entries represent numerical estimates for this setup rather than rigorous error bounds.}
\label{tab:energycheck}
\end{table}

In Bondi gauge, the resulting jump in the asymptotic shear has components
\begin{equation}
 \Delta C_{\theta\theta}=-3a_{\rm mem}\sin^2\theta\,,
 \qquad\Delta C_{\phi\phi}=3a_{\rm mem}\sin^4\theta\,,\label{eq:shearcomponents}
\end{equation}
with $\Delta C_{\theta\phi}=0$, and is generated by the asymptotic supertranslation vector
\begin{equation}
 \xi_f=a_{\rm mem}P_2(\cos\theta)(\partial_u-3\partial_R)
        +\frac{3a_{\rm mem}}R\sin\theta\cos\theta\,\partial_\theta
        +\cdots\,.\label{eq:bmsvector}
\end{equation}
These expressions characterize the relative shift between the early and late endpoint shear in the chosen incoming frame, which should be distinguished from the bulk coordinate transformations that relate the RW, FG, and Bondi descriptions of a fixed local gravitational response.

\subsection{Shear and news}

While the integrated flux fixes the asymptotic memory offset, determining the time-dependent waveform requires evolving the sourced gravitational master field alongside the scalar background. Working in physical coordinates $x = L\rho$, it is convenient to introduce the rescaled time derivative of the master variable, $q = \partial_t p / (8\pi G_4 g\varepsilon)$. Differentiating the sourced master equation eliminates the nonlocal retarded-time primitives, yielding a purely local wave equation:
\begin{equation}
\left(\partial_t^2 - \partial_x^2 + \frac{6F}{R^2}\right)q = -s_t\,, \qquad F = 1 + \frac{R^2}{L^2}\,,
\label{eq:news_evolution_main}
\end{equation}
where the explicit scalar source $s_t$ is given in appendix \ref{app:waveform}. Because eq. \eqref{eq:news_evolution_main} follows directly from the sourced Einstein equations, evolving $q$ from retarded initial data determines the news dynamically without assuming a profile for the matter flux, and the integration constant needed to reconstruct $p$ is fixed by the initial conditions.

Taking the flat-space limit of the initial event and transforming from RW to Bondi gauge, we obtain the asymptotic shear $C_{AB}$ and news tensor $N_{AB}$:
\begin{align}
C_{AB} &= c_2(u) Y_{AB}\,, \qquad c_2(u) = \lim_{R\to\infty} p(u,R)\,,\nonumber\\
N_{AB} &= 8\pi G_4 g\varepsilon \lim_{R\to\infty} \lim_{L\to\infty} q(u,R)\, Y_{AB}\,.
\label{eq:waveform_main}
\end{align}
Here, the inner flat-space limit ($L\to\infty$) is taken at fixed physical extraction radius $R$ and retarded time $u = t - x$, and the choice of incoming soft frame sets $c_2(-\infty) = 0$. 

\begin{samepage}
Specializing to the axisymmetric quadrupole $Y = P_2(\cos\theta)$, the leading TT strain components are $h_+ = 3c_2\sin^2\theta/(2R)$ and $h_\times = 0$, with the corresponding news components obtained by replacing $c_2$ with $\partial_u c_2$.
\end{samepage}

\begin{figure}[tb]
\centering
\includegraphics[width=\textwidth]{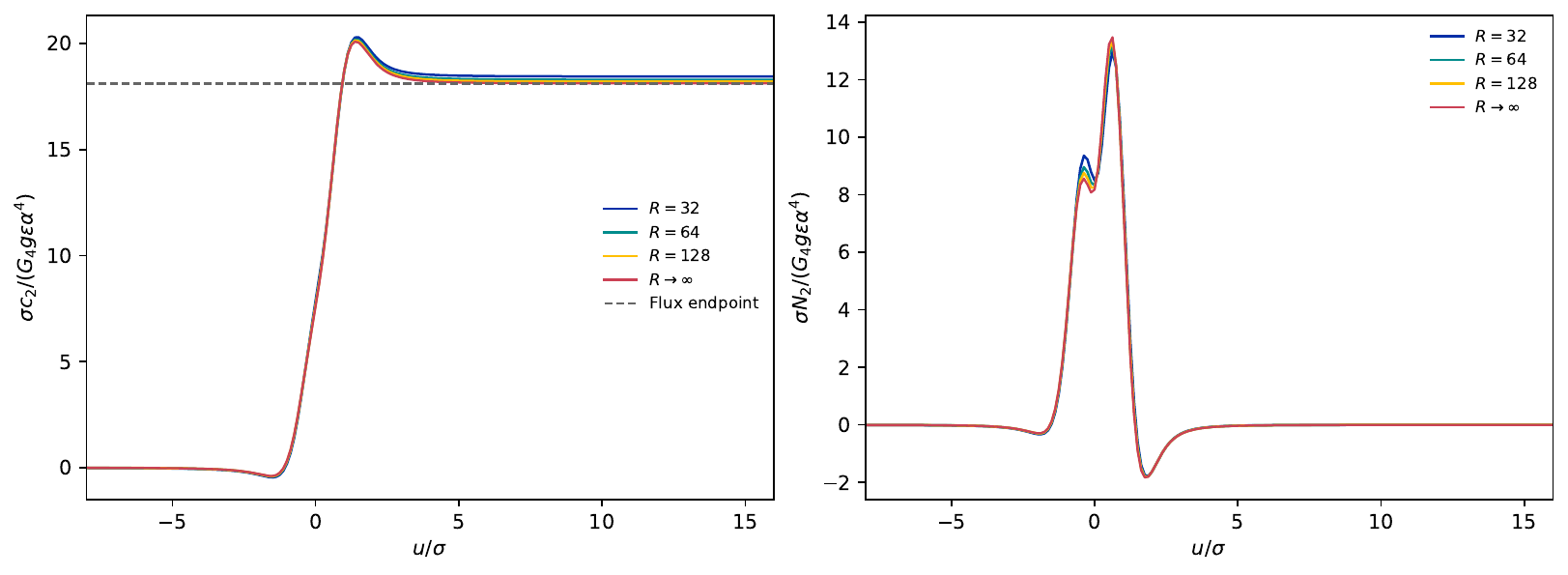}
\caption{Asymptotic shear (left) and news (right) extracted from the sourced master evolution for $\alpha=\sigma=1$. The finitie-$L$ curves show waveforms at finite physical extraction radii $R$ after extrapolating in $L^{-2}$, while the red curves include the subsequent $1/R$ extrapolation to $\mathcal{I}^+$. The dashed horizontal line (left) marks the memory endpoint predicted independently by the integrated matter flux. The numerical settings, window dependence, and convergence tests are given in appendix \ref{app:wavenumerics}.}
\label{fig:waveform}
\end{figure}

Figure \ref{fig:waveform} shows the shear rising, overshooting, and then settling towards a nonzero value, with a short interval of negative news after the main pulse. The extracted jump is consistent with the independent flux calculation:
\begin{equation}
 \frac{\sigma\,[c_2(40\sigma)-c_2(-40\sigma)]}
 {G_4g\varepsilon\alpha^4}\simeq18.12,
 \qquad
 \frac{4\pi}{3}\frac{\sigma\Delta\mathcal E_2}{g\varepsilon\alpha^4}
 \simeq18.12\,.
 \label{eq:wave_memory_number}
\end{equation}
The central values round to the same displayed value for this observation window. Gravitational tails and radial extrapolation limit the comparison,
as discussed in appendix \ref{app:wavenumerics}. The associated supertranslation is $f=-\Delta c_2\,Y/2$.

Beyond the leading radiative data, the radial expansion of the master field, $p = p_0(u) + p_1(u)/R + \cdots$, encodes subleading Bondi quantities. In particular, the quadrupole of the Bondi mass aspect is given by $m_2 = \partial_u p_1$. Comparing $m_2$ extracted directly from the $1/R$ falloff against the integrated constraint equation,
\begin{equation}
m_2(u) = 3p_0(u) - 4\pi G_4 g\varepsilon \int_{-\infty}^u \Phi_2(s)\,d s\,,
\end{equation}
where $\Phi_2$ is the outgoing scalar flux, provides a nontrivial dynamical check throughout the scattering process, during which the mass aspect is nonvanishing. We detail this derivation and confirm its numerical agreement in appendix \ref{app:waveform}.

\section{Discussion}
\label{sec:discussion}
In this work, we have combined the normalizable vacuum mode reconstruction of \cite{LWYGlobalReconstruction} with an explicit sourced reduction to analyze a concrete dynamical Einstein problem. Separating the sourced and homogeneous sectors is essential because the full holographic stress tensor includes the response to bulk matter. While the scalar boundary sources are active, it also generally obeys inhomogeneous Ward identities. Only after subtracting a particular solution generated by the exact same bulk matter history does the boundary response difference yield a conserved, traceless datum $\widehat t_{ij}$ eligible for the global spectral reconstruction. Alongside this global framework, our mode-level check in section \ref{sec:kernel} confirms that the cylindrical boundary integral reproduces the RW master field up to a bulk gauge transformation, without assuming that a purely cut-supported kernel reconstructs the full Fefferman--Graham metric. In the sourced sector, the additional matter-dependent shifts in the RW metric map are dictated directly by the Einstein constraints and must be included independently of the vacuum kernel.

Our interacting scalar scattering model illustrates how these ingredients work together in a dynamical setting. External scalar sources prepare the incoming state, while the bulk fields continue to interact when those sources are switched off. The boundary stress then obeys homogeneous Ward identities, although bulk matter continues to source the gravitational field. Evolving the retarded sourced master equation determines the time-dependent shear and news rather than inferring them from an asymptotic memory constraint. In the ordered flat-space limit, the later boundary operations do not modify the waveform of the first event. The resulting shear offset agrees both with an independent calculation of the outgoing energy redistribution and with the quadrupole moment of the boundary work. Evaluating that work requires the complete sourced stress tensor, including its scalar contact term, rather than the subtracted transverse-traceless tensor $\widehat t_{ij}$ used in homogeneous reconstruction. The transformation used to extract Bondi shear is distinct from the boundary-anchored Fefferman--Graham improvement in the vacuum reconstruction.

While our analysis is restricted to a perturbative scalar scattering process and its electric quadrupole gravitational response, it points toward several natural extensions. Because the homogeneous spectral reconstruction of \cite{LWYGlobalReconstruction} already encompasses general scalar- and vector-type vacuum modes, extending our sourced master reduction to higher multipoles and magnetic parity is a natural next step. Meanwhile, our two-event normal-mode cancellation at finite $L$, the leading $\mathcal{O}(G_4 g\varepsilon)$ waveform, and the ordered flat-space limit used to extract the memory offset are established for a specific smooth scattering family rather than as a uniform theorem over arbitrary source. Developing a systematic analytic treatment of late-time tails and incorporating nonlinear gravitational self-interactions remain important directions for future work. Moreover, our flat-space analysis is restricted to the massless scalar limit obtained at fixed $\Delta=2$. Extending the construction to massive scattering would require a large-$\Delta$ limit with fixed physical mass and a corresponding treatment of timelike matter flux and gravitational memory.

\appendix
\section{Quadrupole integral and master-field consistency}
\label{app:intcheck}

We check eq. \eqref{eq:intmastercheck} without reproducing the seven angular coefficient tables. Throughout this appendix $L=1$, and we
normalize the boundary tensor by $t_{\tau\tau}=Ye^{-\ii\omega\tau}$ with $Y=Y_{2m}$. Its other components are fixed by eq. \eqref{eq:ttmode} with
$a=1/4$. Put $s=\sin\rho$, $c=\cos\rho$, $\Delta=\tau-\tau'$ and $\mu=\hat n\cdot\hat n'$. We denote the cylinder integral by $\mathcal K$. It is defined by
\begin{align}
 \mathcal K[t]_{\mu\nu}
 &=-\frac{3}{8\pi}\int_{S^2}d\Omega'
     \int_{-\alpha(\mu)}^{\alpha(\mu)}d\Delta\,
 \bigl(K_{\mu i}K_{\nu j}-P_{ij}u_\mu u_\nu\bigr)
 t^{ij}(\tau-\Delta,\Omega'),\nonumber\\
 U&=\frac{\cos\Delta-s\mu}{c},\qquad
 u_\mu=\partial_\mu U,\qquad
 K_{\mu i}=\partial_\mu\partial'_iU,\qquad
 \alpha(\mu)=\arccos(s\mu)\,.
 \label{eq:intkernel}
\end{align}
Here $P_{ij}=R_{ij}[\gamma]-R[\gamma]\gamma_{ij}/4$ is the boundary Schouten tensor. The time interval selects the connected spacelike region between the nearest past and future light-cone cuts on the universal cover. This integral defines a pre-FG metric representative, not the boundary-anchored global FG kernel.

Writing $h^{\rm int}_{\mu\nu}=\mathcal K[t]_{\mu\nu}$, we define its seven radial coefficients by
\begin{align}
 (h^{\rm int}_{\tau\tau},h^{\rm int}_{\tau\rho},h^{\rm int}_{\rho\rho})
 &=e^{-\ii\omega\tau}(H_0,H_1,H_2)Y,\nonumber\\
 (h^{\rm int}_{\tau A},h^{\rm int}_{\rho A})
 &=e^{-\ii\omega\tau}(H_3,H_4)D_A Y,\nonumber\\
 h^{\rm int}_{AB}
 &=e^{-\ii\omega\tau}\left(H_5q_{AB}Y+H_6Y_{AB}\right)\,.
 \label{eq:intseven}
\end{align}
The central interval $|\Delta|<q_-=\pi/2-\rho$ contains the full sphere. In the two outer intervals $q_-<|\Delta|<q_+=\pi/2+\rho$, the angular domain is $-1\leq\mu\leq\chi$ with $\chi=\cos\Delta/s$. The angular integrations reduce to the cap moments and the rim value $P_2(\chi)$, with
\begin{equation}
 M_j(\chi)=\int_{-1}^{\chi}\mu^jP_2(\mu)d\mu,
 \qquad j=0,1,2\,.
 \label{eq:intmoments}
\end{equation}
After inserting the derivatives of $U$ and projecting onto the harmonics in eq. \eqref{eq:intseven}, the time integrands are finite Laurent polynomials in $e^{\ii\Delta}$. Each term can be integrated using
\begin{equation}
 {\mathfrak J}_\nu(a,b)=
 \begin{cases}
 \dfrac{e^{\ii\nu b}-e^{\ii\nu a}}{\ii\nu},&\nu\ne0,\\[3pt]
 b-a,&\nu=0
 \end{cases}\,.
 \label{eq:intendpoints}
\end{equation}
The second case is required for $\omega=3,5$. Omitting it would lose the constant Fourier terms in the two outer intervals.

To compare with the RW metric, write
\begin{align}
 \eta&=e^{-\ii\omega\tau}\left(
       -c^2TY\partial_\tau+c^2R_\eta Y\partial_\rho
       +V D^AY\partial_A\right),\nonumber\\
 V&=-\frac{H_6}{2r^2},\qquad
 T=-H_3+\ii\omega r^2V,\qquad
 R_\eta=-H_4-r^2\partial_\rho V\,.
 \label{eq:intgauge}
\end{align}
These relations remove the traceless angular, temporal-angular and radial-angular components in $h^{\rm RW}=\mathcal K[t]+{\cal L}_\eta\bar g$.
Using the transformed coefficients $B_{\rm RW}$ and $K_{\rm RW}$, we obtain
\begin{align}
 p_{\rm int}(\rho)
 &=\frac{\sin(\pi\omega/2)}{4(1-\omega^2)}
 \left[(1-\omega^2+3\cot^2\rho)\sin(\omega\rho)
          -3\omega\cot\rho\cos(\omega\rho)\right]
 \nonumber\\
 &=\frac14e_n^{(b)}(\rho),\qquad \omega=3+2n\,.
 \label{eq:intclosed}
\end{align}
For example, the first two modes give $p_{\rm int}=\sin^3\rho/4$ and $p_{\rm int}=\sin^3\rho(1-8\cos^2\rho)/4$. The remaining components obey
\begin{align}
 K_{\rm RW}&=p_{\rm int}'+\frac{3p_{\rm int}}{r},&
 B_{\rm RW}&=-\frac{\ii\omega}{F}(rp_{\rm int}'+p_{\rm int}),\nonumber\\
 A_{\rm RW}&=p_{\rm int}'+\left(\frac{3F}{r}-r\omega^2\right)p_{\rm int},&
 h^{\rm RW}_{rr}&=\frac{A_{\rm RW}}{F^2}Y e^{-\ii\omega\tau}\,.
 \label{eq:intRWresiduals}
\end{align}
where primes denote $\rho$ derivatives. Thus the integral is gauge-equivalent to the regular RW vacuum mode with the same boundary normalization for every $\omega=3+2n$. Negative-frequency modes follow by complex conjugation. The vector in eq. \eqref{eq:intgauge} need not vanish at the boundary. This calculation establishes a mode-level equality up to a diffeomorphism and makes no claim that this integral alone is the complete boundary-anchored FG reconstruction of
\cite{LWYGlobalReconstruction}.

\section{Source reduction at nonzero and zero frequency}
\label{app:source}
In this appendix $L=1$, and primes on source functions mean $r$ derivatives. For $\omega\ne0$, the auxiliary variable $p=\tfrac12(rK-\ii FB/\omega)$ reduces the equations for the source defined in section \ref{sec:source} to eq. \eqref{eq:sourcedmaster}, with
\begin{align}
 \kappa&=\frac{\ii F}{2\omega}(rR_s+2U_s),\nonumber\\
 \alpha_s&=F\left[rFV_s-Q_s+\frac{\ii}{2\omega}
 \{r^2FR_s'+r(5r^2+3)R_s-4U_s\}\right],\nonumber\\
 \mathcal S_{\rm eff}&=F^2\left[V_s+\frac{\ii}{\omega}(R_s-U_s')\right]
             -\frac{2\ii rF}{\omega}U_s-\frac FrQ_s\,.
 \label{eq:sourcefunctions}
\end{align}
The angular traceless equation sets $h_{rr}=A/F^2$ because its source vanishes. After imposing the master equation \eqref{eq:sourcedmaster}, substituting eqs. \eqref{eq:Dmap}, \eqref{eq:sourcecorrections} and \eqref{eq:sourcefunctions} into the remaining six component equations leaves only the three trace-reversed conservation identities and their derivatives. This supplies a direct check of the reduction against the covariant Einstein equations, independent of the Green function.

For clarity, the actual scalar source can also be written before Fourier transformation. Let
$\phi_I=U_0+\varepsilon P_0Y+g(U_1+\varepsilon P_1Y)$ for both species, and define $W_s=-4(U_0P_1+U_1P_0)+U_0^3P_0$. After removing the common factor $8\pi G_4g\varepsilon$, its components are
\begin{align}
 \calR_{\tau\tau}/Y&=4(U_{0,\tau}P_{1,\tau}+U_{1,\tau}P_{0,\tau})-FW_s,\nonumber\\
 R_s&=2(U_{0,\tau}P_{1,r}+U_{0,r}P_{1,\tau}
       +U_{1,\tau}P_{0,r}+U_{1,r}P_{0,\tau}),\nonumber\\
 \calR_{rr}/Y&=4(U_{0,r}P_{1,r}+U_{1,r}P_{0,r})+W_s/F,\nonumber\\
 U_s&=2(U_{0,\tau}P_1+U_{1,\tau}P_0),\qquad
 V_s=2(U_{0,r}P_1+U_{1,r}P_0),\qquad Q_s=r^2W_s\,.
 \label{eq:actualsource}
\end{align}
Fourier transforms apply to these products, rather than to their factors separately. Scalar equations imply the source conservation identities at the retained order.

The two homogeneous radial solutions used in eq. \eqref{eq:green} are
\begin{align}
 u_\omega(\rho)&=\frac{15[(1-\omega^2+3\cot^2\rho)\sin\omega\rho
                  -3\omega\cot\rho\cos\omega\rho]}
                  {\omega(\omega^2-1)(\omega^2-4)},\nonumber\\
 v_\omega(\rho)&=\frac{(1-\omega^2+3\cot^2\rho)\cos[\omega(\rho-b)]
                  +3\omega\cot\rho\sin[\omega(\rho-b)]}{1-\omega^2}\,.
 \label{eq:uv}
\end{align}
Removable singularities at $\omega=0,\pm1,\pm2$ are defined by analytic continuation. They are not additional gravitational normal modes. For the $g\varepsilon$ source considered here, the Born boundary value vanishes during the free pulse, and the later extraction does not overlap with it. Hence $\delta\sum_IJ_I^2=0$ at this order, and the associated master boundary condition is homogeneous Neumann. This explains the boundary condition used for this source in the main text.

The $1/\omega$ terms in eq. \eqref{eq:sourcefunctions} cannot be ignored when discussing endpoints. Let $R_0=R_s(0,r)$ and $U_0=U_s(0,r)$ in this paragraph only. Source conservation gives
\begin{equation}
 U_0=\frac{r^2}{6}\left[FR_0'+(4r+2/r)R_0\right]\,.
 \label{eq:zerocons}
\end{equation}
The corresponding residue of the auxiliary master variable is $p_{-1}=-\ii Fr^2R_0/6$ in $p=p_{-1}/(\omega+\ii0)+O(1)$. Inserting this residue in the full metric map cancels the poles of $A$ and $K$. The coefficient $B$ has the finite limit $r^2R_0/3$. Thus a primitive tail in $p$ is not by itself a residual physical metric mode. Finally, a regular static homogeneous solution with $p'(b)=0$ would obey
\begin{equation}
 0=\int_0^b\overline p H_2p\,d\rho
  =\int_0^b\left(|p'|^2+6|p|^2/\sin^2\rho\right)d\rho\,,
\end{equation}
and must vanish. Together with eq. \eqref{eq:echo}, this proves cancellation of the residual physical quadrupole after the two completed, compactly supported source histories.

For reference, the matter contributions to the FG constraints in the counterterm scheme of eq. \eqref{eq:stress} are
\begin{align}
 \delta g_{(2)ij}^{\rm matter}&=-2\pi G_4\sum_IJ_I^2\gamma_{ij},\nonumber\\
 \operatorname{tr}g_{(3)}&=-\frac{32\pi G_4}{3}\sum_IJ_IV_I,\qquad
 \nabla^ig_{(3)ij}=-\frac{16\pi G_4}{3}\sum_IJ_I\partial_jV_I\,.
 \label{eq:FGmatter}
\end{align}
These equations, together with the scalar contact term in $T_{ij}$, recover both Ward identities in section \ref{sec:setup}. The quartic interaction is ultraviolet finite at this order and adds no divergent counterterm. The vacuum FG normalization is inherited from \cite{LWYGlobalReconstruction} and is consistent with the separate mode-level check in appendix \ref{app:intcheck}. Our sourced asymptotic extraction is tested numerically. These are different levels of verification, and the latter should not be interpreted as an analytic proof of every sourced boundary coefficient.

\section{Scalar preparation and the memory integral}
\label{app:matter}
\subsection{Free data and boundary control}
The generators used in section \ref{sec:memory} give the exact finite-$L$ free fields
\begin{align}
 \psi_0^{(0)}&=A_0(u)-A_0(v),\nonumber\\
 \psi_2^{(0)}&=A_2''(u)-A_2''(v)
 +\frac{3\cot\rho}{L}[A_2'(u)+A_2'(v)]
 +\frac{3\cot^2\rho+1}{L^2}[A_2(u)-A_2(v)]\,,
 \label{eq:freefields}
\end{align}
where $u=L(\tau-\rho)$ and $v=L(\tau+\rho)$. Primes now differentiate the physical argument of $A_\ell$. The odd difference and derivative terms enforce central regularity. For zero Born boundary data, the normalized scalar modes are $e_{\ell n}=N_{\ell n}\sin^{\ell+1}\rho\,C_{2n+1}^{(\ell+1)}(\cos\rho)$.
If $F_\ell^B$ is the right side of eq. \eqref{eq:bornads}, the retarded coefficients are
\begin{equation}
 q_{\ell n}(\tau)=\int_{-\infty}^{\tau}d s\,
 \frac{\sin[\omega_{\ell n}(\tau-s)]}{\omega_{\ell n}}
 \int_0^bd\rho\,e_{\ell n}(\rho)F_\ell^B(s,\rho)\,.
 \label{eq:scalarretarded}
\end{equation}

After the free field and Born source vanish, a boundary value $j_\ell(s)$ drives the residual modes according to $\ddot q_n+\omega_n^2q_n=-d_nj_\ell$. Vanishing final position and velocity are equivalent to
\begin{equation}
 d_n\int_0^{2\pi}e^{-\ii\omega_ns}j_\ell(s)d s=p_n+\ii\omega_nq_n\,.
 \label{eq:controlmoment}
\end{equation}
The partition of unity in eq. \eqref{eq:control} implies $\int_0^{2\pi}w(s)e^{2\ii ks}d s=\pi\delta_{k0}$. All scalar frequencies in the $\ell=0,2$ sectors are positive even integers, so substituting eq. \eqref{eq:control} proves eq. \eqref{eq:controlmoment} for every mode. Rapid modal decay ensures a smooth control. A nonzero Dirichlet lift may be chosen as $B_\ell(\rho)=\sin^{\ell+1}\rho$, for which $H_\ell B_\ell=(\ell+1)^2B_\ell$. Including that lift is necessary in a direct evolution with the control.

The complete boundary source is
\begin{equation}
 J_I=\frac1L\left[\psi_0^{(0)}(\tau,b)+\varepsilon\psi_2^{(0)}(\tau,b)Y
                  +g j_0(\tau-T)+g\varepsilon j_2(\tau-T)Y\right]\,,
\end{equation}
with the controls supported after the free pulse. For localized incoming generators, the interval between initial preparation and these later controls has $J_I=0$, although the scalar field in the bulk is generally nonzero. Independent copies of this entire protocol, delayed by $\pi$, produce the gravitational completion described in section \ref{sec:source}.

\subsection{Flat Born flux and Bondi conventions}
Let $\phi_I=U+\varepsilon\Psi Y+g(\delta U+\varepsilon\delta\Psi Y)$ in the flat limit. The equations at first order, with $\Box=-\partial_t^2+\nabla^2$, are
\begin{equation}
 \Box\delta U=\tfrac12U^3,\qquad
 \Box(\delta\Psi Y)=\tfrac32U^2\Psi Y\,.
\end{equation}
For a multipolar source $s_\ell(t,R)Y_{\ell m}$, its outgoing coefficient is
\begin{equation}
 \delta F_\ell(u)=-\frac12\int_0^\infty R^2d R
       \int_{-1}^1d x\,P_\ell(x)s_\ell(u+Rx,R)\,.
 \label{eq:flatretarded}
\end{equation}
This is the ordinary retarded Green function followed by the angular projection. The free outgoing coefficients are $A_0(u)$ and $A_2''(u)$. Summing the fluxes of the two species and keeping the $g\varepsilon$ term gives
\begin{equation}
 \frac{\Delta\mathcal E_2}{g\varepsilon}
 =4\int_{-\infty}^{\infty}d u\,
       [A_0'(u)\delta F_2'(u)+A_2'''(u)\delta F_0'(u)]\,.
 \label{eq:bornflux}
\end{equation}
This integral gives the energy redistribution in eq. \eqref{eq:number}. The mass term disappears in the flat limit, and the contribution of $N_{AB}N^{AB}$ to memory \cite{Christodoulou,Favata} is higher order in $G_4$ than the effect retained here.

The Bondi constraint in our conventions reads
\begin{equation}
 \partial_um_B=\tfrac14D_AD_BN^{AB}-\tfrac18N_{AB}N^{AB}
               -4\pi G_4T_{uu}^{(0)}\,.
 \label{eq:bondiconstraint}
\end{equation}
For $\mathcal D_{AB}=D_AD_B-\tfrac12q_{AB}D^2$,
$D^AD^B\mathcal D_{AB}f=\tfrac12D^2(D^2+2)f$. Integrating eq. \eqref{eq:bondiconstraint} with $\Delta C_{AB}=-2\mathcal D_{AB}f$ proves eq. \eqref{eq:bondimemory}. Since $D^2(D^2+2)P_2=24P_2$, the matched relative flux relation gives eq. \eqref{eq:amemory}. Finally,
$\xi_f=f\partial_u+\tfrac12D^2f\partial_R-R^{-1}D^Af\partial_A+\cdots$ has $\Lie_{\xi_f}\eta_{AB}=-2R\mathcal D_{AB}f+\cdots$, fixing the sign of eq. \eqref{eq:bmsvector} directly.

\section{From the sourced master field to a Bondi waveform}
\label{app:waveform}
\subsection{An evolution equation regular at zero frequency}
It is useful to restore physical coordinates before taking a flat limit. Set $x=L\rho$, $R=L\tan(x/L)$ and $F=1+R^2/L^2$, so that $\partial_x=F\partial_R$. The physical master variable is $p=L p_{L=1}$, with the coordinate arguments rescaled at the same time. Its sourced equation is
\begin{align}
 (\partial_x^2-\partial_t^2-6F/R^2)p&=\mathcal S,\nonumber\\
 \mathcal S&=F^2[V_s+\partial_t^{-1}(R_s-\partial_RU_s)]
 -\frac{2RF}{L^2}\partial_t^{-1}U_s-\frac FRQ_s\,.
 \label{eq:physicalmaster}
\end{align}
Here $R_s=\calR_{tR}/Y$, $U_s=\calR_{tA}/D_AY$, $V_s=\calR_{RA}/D_AY$ and $Q_s=\calR_{AB}/(q_{AB}Y)$ are source coefficients in physical coordinates, including $8\pi G_4g\varepsilon$. Every primitive is retarded, with lower endpoint $-\infty$. The metric map becomes
\begin{align}
 K&=Fp_{,R}+3p/R+\kappa,\nonumber\\
 B&=Rp_{,tR}+p_{,t}/F+R\kappa_{,t}/F,\nonumber\\
 A&=Fp_{,R}+3Fp/R+Rp_{,tt}+\alpha_s,\label{eq:physicalmap}\\
 \kappa&=\frac F2\partial_t^{-1}(RR_s+2U_s),\nonumber\\
 \alpha_s&=F\left[RFV_s-Q_s+\frac12\partial_t^{-1}
 \{R^2F\partial_RR_s+R(5R^2/L^2+3)R_s-4U_s\}\right]\,.\nonumber
\end{align}
The physical RW components are $h_{tt}=AY$, $h_{tR}=BY$, $h_{RR}=A Y/F^2$ and $h_{AB}=R^2Kq_{AB}Y$.

Write $z_\ell=\psi_\ell^{(0)}$ and $b_\ell=\psi_\ell^{(1)}$ for the scalar coefficients in eq. \eqref{eq:matterexp}, and define
\begin{equation}
 q=\frac{\partial_t p}{8\pi G_4g\varepsilon}\,.
 \label{eq:qdefinition}
\end{equation}
Differentiating eq. \eqref{eq:physicalmaster}, rather than dividing a Fourier transform by its frequency, gives
\begin{align}
 (\partial_t^2-\partial_x^2+6F/R^2)q&=-s_t,\label{eq:qwave}\\
 s_t&=\frac{4F}{R^2}
 \left[(z_{0,x}-z_0/R)b_{2,t}+(b_{0,x}-b_0/R)z_{2,t}\right]
 -\frac F{R^3}\partial_t(z_0^3z_2)\,.\label{eq:stexplicit}
\end{align}
In particular, this source is determined by the scalar solution and its first derivatives, not by a prescribed news or a target memory.

Here is the cancellation behind eq. \eqref{eq:stexplicit}. Strip the common factor $8\pi G_4g\varepsilon$ from the sources, write the two products as $(U_0,P_1)$ and $(U_1,P_0)$, and put $W=-4\sum UP/L^2+U_0^3P_0$. Direct differentiation gives
\begin{align}
 \frac{\partial_t\mathcal S}{8\pi G_4g\varepsilon}
 &=F^2(V_{s,t}+R_s-U_{s,R})-2RFU_s/L^2-FR W_{,t}\nonumber\\*
 &=4F\sum(FU_{,R}+RU/L^2)P_{,t}-FR\partial_t(U_0^3P_0)\,.
 \label{eq:stcancel}
\end{align}
The mass contribution cancels the unwanted $U_{,t}P$ terms. Substituting $U=z/R$ and $P=b/R$, with $FU_{,R}+RU/L^2=(z_{,x}-z/R)/R$, proves the stated formula. The Born equations are evolved alongside it:
\begin{equation}
 (\partial_t^2-\partial_x^2)b_0=-\frac F{2R^2}z_0^3,
 \qquad
 (\partial_t^2-\partial_x^2+6F/R^2)b_2=-\frac{3F}{2R^2}z_0^2z_2\,.
 \label{eq:physicalborn}
\end{equation}
At the center $b_0=O(x)$ and $b_2,q=O(x^3)$. At $x=\pi L/2$ we impose $b_0=b_2=0$ and $q_{,x}=0$ during the first process. The last condition is the time derivative of the Neumann condition for this source discussed in appendix \ref{app:source}. All three fields have retarded zero initial data.
Recovering $p$ by its retarded time integral then solves the undifferentiated equation too: the possible static discrepancy vanishes on the initial slice. This retains the constraint constant at zero frequency.

\subsection{The asymptotic gauge transformation}
The Bondi statement concerns the flat limit of the first event. In this subsection take $L\to\infty$ first and put $u=t-R$. Suppose the retarded solution has the outgoing expansion
\begin{equation}
 p(u,R)=p_0(u)+\frac{p_1(u)}R+\frac{p_2(u)}{R^2}+\cdots\,.
 \label{eq:pexp}
\end{equation}
We use the usual smooth asymptotic class at the displayed orders. The numerical extrapolation below tests this class for the specified data, rather than proving it uniformly in $L$.
Let $\Phi_2(u)$ be the coefficient of $g\varepsilon Y/R^2$ in the flat outgoing matter flux. Thus
\begin{equation}
 \Phi_2=4[A_0'\delta F_2'+A_2'''\delta F_0'],\qquad
 J(u)=8\pi G_4g\varepsilon\int_{-\infty}^u\Phi_2(s)d s\,.
 \label{eq:waveJ}
\end{equation}
Expanding the actual source gives
\begin{equation}
 \mathcal S=-\frac{J}{R^2}+O(R^{-3}),\quad
 R_s=-\frac{J'}{R^2}+O(R^{-3}),\quad
 \kappa=-\frac{J}{2R}+O(R^{-2}),\quad
 \alpha_s=\frac{J'}2+O(R^{-1})\,.
 \label{eq:asympsource}
\end{equation}
The $R^{-2}$ master equation therefore fixes
\begin{equation}
 p_1'=3p_0-\frac J2\,.
 \label{eq:p1constraint}
\end{equation}
Consequently $K=-p_0'+O(R^{-2})$: its putative $R^{-1}$ coefficient cancels between the homogeneous map and $\kappa$. The same cancellation gives $A+B=O(R^{-2})$. Keeping these source terms is essential to a regular Bondi transformation.

For clarity, the cancellation in $A+B$ does not require an assumption about subleading scalar coefficients. Write $R_s=-J'/R^2+Q_3/R^3+\cdots$, $U_s=U_2/R^2+\cdots$, and $V_s=V_2/R^2+\cdots$. Outgoing matter gives $V_2=-U_2$. If $\kappa=-J/(2R)+\kappa_2/R^2+\cdots$ and $\alpha_s=J'/2+\alpha_1/R+\cdots$, eq. \eqref{eq:physicalmap} yields
\begin{equation}
 \alpha_1=V_2-J/2-Q_3/2,\qquad
 \kappa_2'=(Q_3+2U_2)/2,\qquad
 \alpha_1+\kappa_2'=-J/2\,.
 \label{eq:asympcancellation}
\end{equation}
Together with eq. \eqref{eq:p1constraint}, this removes the $R^{-1}$ term of $A+B$, while $\alpha_0+\lim R\kappa_{,t}=0$ removes its constant term.

Use $h^{\rm B}=h^{\rm RW}+\Lie_{\xi_B}\eta$, where $\xi_B=U_B Y\partial_u+V_B Y\partial_R+W_B D^AY\partial_A$. The radial Bondi conditions and the determinant condition are solved by
\begin{align}
 U_B(u,R)&=-\tfrac12p_0(u)-\int_R^\infty(A+B)(u,s)d s,\nonumber\\
 W_B(u,R)&=-\int_R^\infty U_B(u,s)s^{-2}d s,\nonumber\\
 V_B(u,R)&=3RW_B-\tfrac12RK\,.\label{eq:bondigaugeintegrals}
\end{align}
Indeed $h^{\rm B}_{RR}=2(A+B)-2U_{B,R}=0$, $h^{\rm B}_{RA}=(R^2W_{B,R}-U_B)D_AY=0$ and the angular trace vanishes. The remaining mixed component is $h^{\rm B}_{uR}=-U_{B,u}-V_{B,R}$.
Its leading term fixes $\partial_uU_B(\infty)=-p_0'/2$. The integration constant is fixed by the incoming soft frame. Since $W_B=p_0/(2R)+O(R^{-2})$, the traceless angular metric gives
\begin{align}
 C_{AB}&=p_0Y_{AB},\qquad N_{AB}=p_0'Y_{AB},\nonumber\\
 p_0'&=8\pi G_4g\varepsilon\lim_{R\to\infty}\lim_{L\to\infty}q(t=x+u,x)\,.
 \label{eq:directbondi}
\end{align}
This is a metric extraction, obtained before using a flux constraint to evaluate its endpoint. The vector $\xi_B$ is separate from the vector relating the original holographic integral to RW gauge.

There is a useful additional check. The radial Einstein equation and $T_{RR}=O(R^{-4})$ give $h^{\rm B}_{uR}=O(R^{-2})$. In the expansion $U_B=-p_0/2+u_1/R+\cdots$, this implies $u_1'=0$. The $R^{-1}$ coefficient of $h^{\rm B}_{uu}$, using eq. \eqref{eq:asympcancellation}, is then
$p_1'+3p_0+\alpha_1+\kappa_2'=2p_1'$. Thus the quadrupole mass aspect is
\begin{equation}
 m_2=p_1'=3p_0-J/2,\qquad
 m_2'=3p_0'-4\pi G_4g\varepsilon\Phi_2\,.
 \label{eq:directmass}
\end{equation}
This agrees with the Bondi constraint \eqref{eq:bondiconstraint}: the harmonic satisfies $D^AD^BY_{AB}=12Y$.
Only at the endpoints, with the prescribed incoming matching and no massive remnant, does this reduce to
$\Delta p_0=(4\pi G_4/3)\Delta\mathcal E_2$. The constraint alone would not have determined $p_0(u)$.

\subsection{Numerical extraction and its scope}
\label{app:wavenumerics}
We evolve eqs. \eqref{eq:qwave} and \eqref{eq:physicalborn} together with the analytic finite-$L$ free fields in eq. \eqref{eq:freefields}. The evolution uses fourth-order centered radial differences and fourth-order Runge--Kutta time stepping with $\Delta t\leq0.35\Delta x$. Regular central extensions and the prescribed Dirichlet and Neumann outer conditions are imposed. The Gaussian initial data are set at $t=-\pi L/2-10\sigma$. The omitted tails are negligible at the quoted precision. The extraction windows close before the first boundary reflection returns, so the later controls and delayed copy do not affect the waveform.

At fixed retarded time and physical radius, we first extrapolate quadratically in $L^{-2}$ using $L/\sigma=128,256,512$, and then quadratically in $R^{-1}$ using $R/\sigma=32,48,64,96,128$. The leading terms yield the shear and news, while the subleading coefficient tests the mass aspect. Refining the spatial grid from $\Delta x/\sigma=0.02$ to $0.01$ changes the extracted shear and news at a representative radius by at most $1.59\times10^{-4}$ and $4.58\times10^{-4}$ in units of $G_4g\varepsilon$. Varying the radial fit and extraction radii gives further changes at approximately the $10^{-3}$ level or below. These are numerical convergence diagnostics for this scattering process, not uniform error bounds in $L$ or $R$.

The observation window matters because of the gravitational tail: extending $u\in[-8\sigma,8\sigma]$ to $[-40\sigma,40\sigma]$ changes the normalized shear jump from about $18.134$ to $18.1171$, compared with the independent integrated-flux endpoint $18.1166$. The evolving scalar field also gives a quadrupole energy redistribution consistent with the flat retarded calculation. The subleading Bondi mass-aspect check is less precise, agreeing with the constraint in eq. \eqref{eq:directmass} only at the few-parts-in-a-thousand level relative to the transient mass quadrupole. Thus the leading waveform agreement should not be read as a comparably precise extraction of all subleading coefficients.

This evolution provides a sourced waveform for the specified matter process, not a uniform flat-space-limit theorem for arbitrary boundary data. In the purely retarded setup $h_P$ already contains the physical response and the additional homogeneous gravitational datum vanishes. If freely specified homogeneous perturbations are superimposed, their reconstruction is provided by the global spectral normal modes of \cite{LWYGlobalReconstruction}.

\bibliographystyle{jhep}
\bibliography{references}

@article{Maldacena,
    author = "Maldacena, Juan Martin",
    title = "{The Large $N$ limit of superconformal field theories and supergravity}",
    eprint = "hep-th/9711200",
    archivePrefix = "arXiv",
    reportNumber = "HUTP-97-A097, HUTP-98-A097",
    doi = "10.4310/ATMP.1998.v2.n2.a1",
    journal = "Adv. Theor. Math. Phys.",
    volume = "2",
    pages = "231--252",
    year = "1998"
}

@article{GKP,
    author = "Gubser, S. S. and Klebanov, Igor R. and Polyakov, Alexander M.",
    title = "{Gauge theory correlators from noncritical string theory}",
    eprint = "hep-th/9802109",
    archivePrefix = "arXiv",
    reportNumber = "PUPT-1767",
    doi = "10.1016/S0370-2693(98)00377-3",
    journal = "Phys. Lett. B",
    volume = "428",
    pages = "105--114",
    year = "1998"
}

@article{Witten,
    author = "Witten, Edward",
    title = "{Anti de Sitter space and holography}",
    eprint = "hep-th/9802150",
    archivePrefix = "arXiv",
    reportNumber = "IASSNS-HEP-98-15",
    doi = "10.4310/ATMP.1998.v2.n2.a2",
    journal = "Adv. Theor. Math. Phys.",
    volume = "2",
    pages = "253--291",
    year = "1998"
}

@article{BDHM,
    author = "Banks, Tom and Douglas, Michael R. and Horowitz, Gary T. and Martinec, Emil J.",
    title = "{AdS dynamics from conformal field theory}",
    eprint = "hep-th/9808016",
    archivePrefix = "arXiv",
    reportNumber = "NSF-ITP-98-082, EFI-98-30",
    month = "8",
    year = "1998"
}

@article{HKLL,
    author = "Hamilton, Alex and Kabat, Daniel N. and Lifschytz, Gilad and Lowe, David A.",
    title = "{Holographic representation of local bulk operators}",
    eprint = "hep-th/0606141",
    archivePrefix = "arXiv",
    reportNumber = "CU-TP-1149",
    doi = "10.1103/PhysRevD.74.066009",
    journal = "Phys. Rev. D",
    volume = "74",
    pages = "066009",
    year = "2006"
}

@article{HKLLcomplex,
    author = "Hamilton, Alex and Kabat, Daniel N. and Lifschytz, Gilad and Lowe, David A.",
    title = "{Local bulk operators in AdS/CFT: A Holographic description of the black hole interior}",
    eprint = "hep-th/0612053",
    archivePrefix = "arXiv",
    reportNumber = "CU-TP-1162",
    doi = "10.1103/PhysRevD.75.106001",
    journal = "Phys. Rev. D",
    volume = "75",
    pages = "106001",
    year = "2007",
    note = "[Erratum: Phys.Rev.D 75, 129902 (2007)]"
}

@article{KLL,
    author = "Kabat, Daniel and Lifschytz, Gilad and Lowe, David A.",
    title = "{Constructing local bulk observables in interacting AdS/CFT}",
    eprint = "1102.2910",
    archivePrefix = "arXiv",
    primaryClass = "hep-th",
    doi = "10.1103/PhysRevD.83.106009",
    journal = "Phys. Rev. D",
    volume = "83",
    pages = "106009",
    year = "2011"
}

@article{DonnellyGiddings,
    author = "Donnelly, William and Giddings, Steven B.",
    title = "{Observables, gravitational dressing, and obstructions to locality and subsystems}",
    eprint = "1607.01025",
    archivePrefix = "arXiv",
    primaryClass = "hep-th",
    doi = "10.1103/PhysRevD.94.104038",
    journal = "Phys. Rev. D",
    volume = "94",
    number = "10",
    pages = "104038",
    year = "2016"
}

@article{LWY,
    author = "Lowe, David A. and Wang, Yiru and Yang, Juanyi",
    title = "{Holographic reconstruction of gravitational perturbations in AdS/CFT and implications for celestial conformal field theory}",
    eprint = "2411.02364",
    archivePrefix = "arXiv",
    primaryClass = "hep-th",
    doi = "10.1007/JHEP04(2025)070",
    journal = "JHEP",
    volume = "04",
    pages = "070",
    year = "2025"
}

@article{LWYGlobalReconstruction,
    author = "Lowe, David A. and Wang, Yiru and Yang, Juanyi",
    title = "{Gravitational Reconstruction in $AdS_{4}$}",
    eprint = "2610.38887",
    archivePrefix = "arXiv",
    primaryClass = "hep-th",
    year = "2026"
}

@article{ReggeWheeler,
    author = "Regge, Tullio and Wheeler, John A.",
    title = "{Stability of a Schwarzschild singularity}",
    doi = "10.1103/PhysRev.108.1063",
    journal = "Phys. Rev.",
    volume = "108",
    pages = "1063--1069",
    year = "1957"
}

@article{Zerilli,
    author = "Zerilli, Frank J.",
    title = "{Effective potential for even parity Regge-Wheeler gravitational perturbation equations}",
    doi = "10.1103/PhysRevLett.24.737",
    journal = "Phys. Rev. Lett.",
    volume = "24",
    pages = "737--738",
    year = "1970"
}

@article{Moncrief,
    author = "Moncrief, V.",
    title = "{Gravitational perturbations of spherically symmetric systems. I. The exterior problem.}",
    doi = "10.1016/0003-4916(74)90173-0",
    journal = "Annals Phys.",
    volume = "88",
    pages = "323--342",
    year = "1974"
}

@article{KI,
    author = "Kodama, Hideo and Ishibashi, Akihiro",
    title = "{A Master equation for gravitational perturbations of maximally symmetric black holes in higher dimensions}",
    eprint = "hep-th/0305147",
    archivePrefix = "arXiv",
    doi = "10.1143/PTP.110.701",
    journal = "Prog. Theor. Phys.",
    volume = "110",
    pages = "701--722",
    year = "2003"
}

@article{IW,
    author = "Ishibashi, Akihiro and Wald, Robert M.",
    title = "{Dynamics in nonglobally hyperbolic static space-times. 3. Anti-de Sitter space-time}",
    eprint = "hep-th/0402184",
    archivePrefix = "arXiv",
    doi = "10.1088/0264-9381/21/12/012",
    journal = "Class. Quant. Grav.",
    volume = "21",
    pages = "2981--3014",
    year = "2004"
}

@article{DiasSantos,
    author = "Dias, {\'O}scar J. C. and Santos, Jorge E.",
    title = "{Boundary Conditions for Kerr-AdS Perturbations}",
    eprint = "1302.1580",
    archivePrefix = "arXiv",
    primaryClass = "hep-th",
    doi = "10.1007/JHEP10(2013)156",
    journal = "JHEP",
    volume = "10",
    pages = "156",
    year = "2013"
}

@article{ZeldovichPolnarev,
    author = "Zel'dovich, Y. B. and Polnarev, A. G.",
    title = "{Radiation of gravitational waves by a cluster of superdense stars}",
    journal = "Sov. Astron.",
    volume = "18",
    pages = "17",
    year = "1974"
}

@article{Christodoulou,
    author = "Christodoulou, D.",
    title = "{Nonlinear nature of gravitation and gravitational wave experiments}",
    doi = "10.1103/PhysRevLett.67.1486",
    journal = "Phys. Rev. Lett.",
    volume = "67",
    pages = "1486--1489",
    year = "1991"
}

@article{Favata,
    author = "Favata, Marc",
    title = "{The gravitational-wave memory effect}",
    eprint = "1003.3486",
    archivePrefix = "arXiv",
    primaryClass = "gr-qc",
    doi = "10.1088/0264-9381/27/8/084036",
    journal = "Class. Quant. Grav.",
    volume = "27",
    pages = "084036",
    year = "2010"
}

@article{Bondi,
    author = "Bondi, H. and van der Burg, M. G. J. and Metzner, A. W. K.",
    title = "{Gravitational waves in general relativity. 7. Waves from axisymmetric isolated systems}",
    doi = "10.1098/rspa.1962.0161",
    journal = "Proc. Roy. Soc. Lond. A",
    volume = "269",
    pages = "21--52",
    year = "1962"
}

@article{Sachs,
    author = "Sachs, R. K.",
    title = "{Gravitational waves in general relativity. 8. Waves in asymptotically flat space-times}",
    doi = "10.1098/rspa.1962.0206",
    journal = "Proc. Roy. Soc. Lond. A",
    volume = "270",
    pages = "103--126",
    year = "1962"
}

@article{Weinberg,
    author = "Weinberg, Steven",
    title = "{Infrared photons and gravitons}",
    doi = "10.1103/PhysRev.140.B516",
    journal = "Phys. Rev.",
    volume = "140",
    pages = "B516--B524",
    year = "1965"
}

@article{Strominger,
    author = "Strominger, Andrew",
    title = "{On BMS Invariance of Gravitational Scattering}",
    eprint = "1312.2229",
    archivePrefix = "arXiv",
    primaryClass = "hep-th",
    doi = "10.1007/JHEP07(2014)152",
    journal = "JHEP",
    volume = "07",
    pages = "152",
    year = "2014"
}

@article{HeLysovMitraStrominger,
    author = "He, Temple and Lysov, Vyacheslav and Mitra, Prahar and Strominger, Andrew",
    title = "{BMS supertranslations and Weinberg{\textquoteright}s soft graviton theorem}",
    eprint = "1401.7026",
    archivePrefix = "arXiv",
    primaryClass = "hep-th",
    doi = "10.1007/JHEP05(2015)151",
    journal = "JHEP",
    volume = "05",
    pages = "151",
    year = "2015"
}

@article{SZ,
    author = "Strominger, Andrew and Zhiboedov, Alexander",
    title = "{Gravitational Memory, BMS Supertranslations and Soft Theorems}",
    eprint = "1411.5745",
    archivePrefix = "arXiv",
    primaryClass = "hep-th",
    doi = "10.1007/JHEP01(2016)086",
    journal = "JHEP",
    volume = "01",
    pages = "086",
    year = "2016"
}

@article{LoweRamirez,
    author = "Lowe, David A. and Ramirez, David M.",
    title = "{BMS Symmetry via AdS/CFT}",
    eprint = "2007.02851",
    archivePrefix = "arXiv",
    primaryClass = "hep-th",
    doi = "10.1007/JHEP01(2021)075",
    journal = "JHEP",
    volume = "01",
    pages = "075",
    year = "2021"
}

@article{CompereFiorucciRuzziconi,
    author = "Comp{\`e}re, Geoffrey and Fiorucci, Adrien and Ruzziconi, Romain",
    title = "{The $\Lambda$-BMS$_4$ group of dS$_4$ and new boundary conditions for AdS$_4$}",
    eprint = "1905.00971",
    archivePrefix = "arXiv",
    primaryClass = "gr-qc",
    doi = "10.1088/1361-6382/ab3d4b",
    journal = "Class. Quant. Grav.",
    volume = "36",
    number = "19",
    pages = "195017",
    year = "2019",
    note = "[Erratum: Class.Quant.Grav. 38, 229501 (2021)]"
}

@article{BalasubramanianKraus,
    author = "Balasubramanian, Vijay and Kraus, Per",
    title = "{A Stress tensor for Anti-de Sitter gravity}",
    eprint = "hep-th/9902121",
    archivePrefix = "arXiv",
    reportNumber = "HUTP-99-A002, EFI-99-6, NSF-ITP-98-132",
    doi = "10.1007/s002200050764",
    journal = "Commun. Math. Phys.",
    volume = "208",
    pages = "413--428",
    year = "1999"
}

@article{deHaro,
    author = "de Haro, Sebastian and Solodukhin, Sergey N. and Skenderis, Kostas",
    title = "{Holographic reconstruction of space-time and renormalization in the AdS / CFT correspondence}",
    eprint = "hep-th/0002230",
    archivePrefix = "arXiv",
    reportNumber = "SPIN-2000-05, ITP-UU-00-03, PUTP-1921",
    doi = "10.1007/s002200100381",
    journal = "Commun. Math. Phys.",
    volume = "217",
    pages = "595--622",
    year = "2001"
}

@article{BianchiFreedmanSkenderis,
    author = "Bianchi, Massimo and Freedman, Daniel Z. and Skenderis, Kostas",
    title = "{Holographic renormalization}",
    eprint = "hep-th/0112119",
    archivePrefix = "arXiv",
    reportNumber = "MIT-CTP-3166, PUTP-1999, DAMTP-2001-63, ROM2F-2001-30",
    doi = "10.1016/S0550-3213(02)00179-7",
    journal = "Nucl. Phys. B",
    volume = "631",
    pages = "159--194",
    year = "2002"
}

@article{Skenderis,
    author = "Skenderis, Kostas",
    title = "{Lecture notes on holographic renormalization}",
    eprint = "hep-th/0209067",
    archivePrefix = "arXiv",
    reportNumber = "PUTP-2047",
    doi = "10.1088/0264-9381/19/22/306",
    journal = "Class. Quant. Grav.",
    volume = "19",
    pages = "5849--5876",
    year = "2002"
}
\end{document}